\documentclass[preprintnumbers,eqsecnum,aps,prd,epsf,showpacs,nofootinbib]{revtex4}
\usepackage{latexsym,amsmath,amssymb}
\usepackage[dvips]{graphicx,color}% Include figure files
\usepackage{dcolumn,bm}    % Align table columns on decimal point
\usepackage{subfigure}  % sub labeling of fingures
\usepackage{color}
\begin{document}
\thispagestyle{empty}
\title{The effect of varying sound velocity 
on primordial curvature 
perturbations}

\preprint{RESCEU-22-10}

\author{Masahiro Nakashima $^{1,2}$}
\email{nakashima_at_resceu.s.u-tokyo.ac.jp}

\author{Ryo Saito $^{1,2}$}
\email{r-saito_at_resceu.s.u-tokyo.ac.jp}

\author{Yu-ichi Takamizu$^{2}$}
\email{takamizu_at_resceu.s.u-tokyo.ac.jp}

\author{Jun'ichi Yokoyama$^{2,3}$}
\email{yokoyama_at_resceu.s.u-tokyo.ac.jp}
\affiliation{
\\
$^{1}$ Department of Physics, Graduate School of Science,  
The University of Tokyo, Tokyo 113-0033, Japan \\
$^{2}$ Research Center for the Early Universe (RESCEU), Graduate
School of Science, The University of Tokyo, Tokyo 113-0033, Japan 
\\
$^{3}$ Institute for the Physics and Mathematics of the Universe (IPMU),
The University of Tokyo, 5-1-5 Kashiwanoha, Kashiwa, Chiba 277-8582, Japan
}
\date{\today}

\begin{abstract}
We study the effects of sudden change in the sound velocity on primordial
curvature perturbation spectrum in inflationary cosmology, assuming that
the background evolution satisfies the slow-roll condition throughout.
It is found that the power spectrum acquires oscillating features which are
 determined by the ratio of the
sound speed before and after the transition and the wavenumeber
which crosses the sound horizon
at the transition, and their analytic expression is given.
In some values of those parameters, the
oscillating primordial power spectrum
can better fit the observed Cosmic Microwave Background
temperature anisotropy power spectrum
than the simple power-law power spectrum, although introduction of
such a new degree of freedom is not justified in the context of
Akaike's Information Criterion.

%We study the effect of varying sound velocity on the primordial curvature perturbation. If the inflaton action has non-canonical kinetic terms and those terms couple to some ohter fields, then the sound speed of the perturbation can change during inflation. We find that if we assume the background evolution satisfies the slow-roll condition and the change of the sound speed occurs suddenly then the power spectrum have oscillating features which is determined by the ratio of the sound speed before and after the transition and the wavenumeber which crosses the sound horizon at the transition. In some values of those parameters, the oscillating primordial power spectrum can better fit the obserbed Cosmic Microwave Background temperature anisotropy power spectrum than the simple power-law power spctrum.  
\end{abstract}
\maketitle

\section{Introduction}
\label{sec:intro}
Standard inflationary cosmology \cite{Guth:1980zm,Starobinsky:1980te,Sato:1980yn} predicts nearly scale-invariant power spectrum of the primordial perturbation \cite{Mukhanov:1981xt,Hawking:1982cz,Starobinsky:1982ee,Guth:1982ec}. Such a power-law like perturbation spectrum $\Delta_{\zeta} ^{2} \propto k^{n_{s}-1}$, where $\zeta$ is the comoving curvature perturbation and $n_{s}$ is the scalar spectral index with $n_{s} \simeq 1$ has been prefered also from a number of obserbations \cite{Komatsu:2010fb}. If we look into the detailed structure of the Cosmic Microwave Background (CMB) temperature anisotropy, some hints of small deviations from the simplest form of the curvature perturbation power spectrum show up. Among those, anomalously low values of the quadrupole moment or several sharp glitches in the large scale WMAP data corresponing to $\ell \sim 20-40$ are famous and most intensively studied \cite{Cline:2003ve}. 
Furthermore, several groups have reported strong evidence of the deviation from a simple power-law type power spectrum by reconstructing the primordial power spectrum from the WMAP data \cite{Matsumiya:2001xj,Matsumiya:2002tx,Kogo:2003yb,Kogo:2004vt,Kogo:2005qi,TocchiniValentini:2004ht,TocchiniValentini:2005ja,Nagata:2008tk,Nagata:2008zj,Nagata:2008zzb}. In particular, Ichiki et al. \cite{Ichiki:2009zz,Ichiki:2009xs} claim that they found an oscillatory modulation localized around the comoving wavenumber $k\simeq 0.009[\mathrm{Mpc}^{-1}]$ ($\ell \simeq 120$) in the power spectrum at 99.995$\%$ confidence level.   

Theoretical calculation for the power spectrum has been also sophisticated recently and some inflation models based on a realistic high-energy physics can generate peculiar features on the curvature perturbation. Those include the so-called Trans-Planckian effect \cite{Martin:2000xs,Martin:2000bv,Easther:2002xe}, particle production due to the coupling between the inflaton and another scalar filed \cite{Romano:2008rr,Barnaby:2009mc,Barnaby:2009dd}, temporal violation of the slow-rolling of the inflaton field \cite{Leach:2000yw,Leach:2001zf,Adams:2001vc,Starobinsky:1992ts,Kaloper:2003nv,Joy:2007na,Joy:2008qd,Battefeld:2010rf}, and some other models \cite{Cai:2008ed,Jackson:2010cw}. 
To go further into the accurate cosmology, it is crucial to study further the details of the primordial perturbations. This leads us to the detailed structure of the inflaton Lagrangian or the true high-energy physics that has realized in our Universe.

In this paper, we concentrate on the role of the sound velocity, which is defined as the propagating speed of the linear perturbation in the next section. In some high-energy physics theories, there appear non-canonical kinetic terms in the Lagrangian and in that case the sound velocity deviates from unity \cite{Garriga:1999vw,ArmendarizPicon:1999rj}. Furthermore, if those kinetic terms of the inflaton field couple to some time-dependent variables which can be seen, for example, in DBI inflation scenario \cite{Silverstein:2003hf}, the values of the sound velocity can change during inflation \cite{Wei:2004xx,Kinney:2007ag,Khoury:2008wj}. Then we have to consider the possibility of some non-trivial dynamics of the perturbation. As a result, new degrees of freedom such as the sound velocity may be observed in the CMB temperature fluctuation and be tested in the future high-precision CMB observations such as PLANCK or CMBpol.

This paper is organized as follows. In the next section, we comment on the background evolution in our scenario and introduce the sound velocity. In sec \ref{basic-eq}, the basic variables and its evolution equation for the curvature perturbation are described. Then we consider the paricular types of the variation of the sound velocity. The first type is a step-like function, which is discussed in Sec. \ref{step-variation}, and the second type is a top-hat type function, which we will study in Sec. \ref{top-hat-variation}. The last section is devoted to the summury and discussion.  

%The recent more accurate observations of the cosmic microwave backgrond motivat%e...  

%The further information obtained from the 
%near future observations 
%such as the PLANCK satellite, 
%is need in order to discriminate 
%many possible inflationary models of them. 

%Local `features' such bumps or splikes of CMB power spectrum are seen at 
%$l\sim 100 ? $ or $l\sim 40$ (Archeops feature). 

%If exist, these local features are thought to 
%be generated by a local dynamics, such as varying sound velocity or ...

\section{Background assumption and sound velocity}
\label{Background}
In the standard single inflaton field with a canonical kinetic term, the sound velocity $c_{s}$ defined by the propagation speed of linear perturbation has the value equal to the speed of light, namely, $c_{s}=1$. This is easily checked by considering the action of the canonical inflaton field, 
\begin{equation}
S=\int d^{4}x \sqrt{-g}\left(\frac{1}{2}R+X+V(\phi)\right),\ \ X=-{1\over 2}g^{\mu\nu}\partial_{\mu}\phi\partial_{\nu}\phi,
\end{equation} 
where $R$ denote the Einstein-Hilbert action and we set the reduced Planck scale to unity ($8\pi G=1$). Expanding the action around the background Friedman-Robertson-Walker (FRW) metric up to second-order, then we can see that the sound velocity which appears as the coefficient of the spatial derivative term is exactly equal to one.

Generalizing the above inflaton action to an arbitary function of the kinetic term $X$,
\begin{equation}
S=\int d^{4}x \sqrt{-g}\left(\frac{1}{2}R+P(X,\phi)\right), \label{eq:general_action}
\end{equation}
the situation changes completely. In this action, expanding around FRW metric up to second-order, we find the sound velocity is given by  
\begin{eqnarray}
c_s^{2}= \frac{P_{X}}{2P_{XX}X+P_X} \label{eq:soundvel},
\end{eqnarray}
where the subscript $X$ represents a derivative with respect to $X$. The specific second-order action for the approapriate perturbation variable, what we call the comoving curvature perturbation, will be presented in the next section. It is clear from (\ref{eq:soundvel}) that, in principle, the sound velocity takes various values corresponding to the functional form $P(X,\phi)$ and the values of $X$ and $\phi$.

In this paper, in order to extract the effects of the change of the sound velocity alone, we study the cases it changes suddenly once or twice during inflation without affecting the background evolution. The sudden change of the sound velocity can be realized, for example, if the Lagrangian has terms like
\begin{equation}
P(X, \phi) \supset f(\phi)X+[1-f(\phi)]X^{2}/\Lambda^{4},\ \ f(\phi) =\frac{1}{1+e^{(\phi-\phi_{0})/d}}, \label{csmodel1}
\end{equation}
or 
\begin{equation}
P(X, \phi) \supset b(\phi)X+X^{2}/\Lambda^{4},\ \ b(\phi) \simeq (\phi-\phi_{0})^{2} + \mathcal{O}\left((\phi-\phi_{0})^{n}\right), \ \ \ (n\ge 3)  \label{csmodel2}
\end{equation}  
where $\phi$ is a scaler field which we assume to be the inflaton and $\Lambda$ is some cut-off scale. In the former case, we find $c_{s}^{2}=1$ when $\phi \ll \phi_{0}$ and $c_{s}^{2} =1/3$ when $\phi \gg \phi_{0}$. Such a transition of the sound velocity due to the motion of the field $\phi$ is approximately described by a step function. In the latter case, $c_{s}^{2}=1/3$ only when $\phi \simeq \phi_{0}$ and at the other values of $\phi$, $c_{s}^{2}=1$. This variation is well approxmated by a top-hat type function. In generic unified theories including string theory, there appears many scalar fields which have non-trivial kinetic terms and it is likely that those kinetic terms naturally couple to the scalar field itself in a complicated way. Here we consider simple tractable examples of posssible deviation from the canonical model as described above. 

If we take these models as they are, the background evolution, in particular the slow-roll parameters, may also be severely modulated unless some fine-tunning is applied. Since the effects of the sudden change of slow-roll parameters 
\begin{equation}
\epsilon= -\frac{\dot{H}}{H^{2}},\ \ \ \eta= \frac{\dot{\epsilon}}{H\epsilon},
\end{equation}
where a dot represents a derivative with respect to the physical time $t$,
have been already studied in the literatures \cite{Starobinsky:1992ts,Kaloper:2003nv}, we concentrate on the cases only the sound velocity changes suddenly as mentioned above. Thus, we impose the slow-roll conditions
\begin{equation}
|\epsilon| \ll 1,\ \ \ |\eta| \ll 1, \label{eq:slow-roll_cond}
\end{equation}
throughout this paper. Such a situation can be also realized in the curvaton scenario as discussed in Appendix A. 

%Partially motivated by the above mentioned Lagrangians, we calculate the power spectrum of the curvature perturbation in case the sound velocity changes suddenly like a step function or a top-hat type function during inflation. When the sound velocity changes according to (\ref{csmodel1}) or (\ref{csmodel2}), the background evolution may be affected significantly at the same time. We, however, analyze the case the background remain intact in order to separate the effect due to the change of sound velocity and concentrate on it\footnote{This situation in which the sudden change of the sound velocity does not affect the background evolution is easily realized in the curvaton scenario, which is commented in Appendix A.}. So, throughout this paper, we assume that with the standard single field inflation, the background evolution satisfies slow-roll conditions, $\epsilon,|\eta|\ll 1$ during inflation, where we define slow-roll parameters as

In the usual calculation of linear perturbation in this type of generalized non-canonical kinetic term inflation models, new parameters which parametrize the time variation of the sound velocity such as
\begin{eqnarray}
\epsilon_s={\dot{c}_s\over c_s H}, \ \ \ \eta_{s} = {\dot{\epsilon}_{s}\over \epsilon_{s} H},
\end{eqnarray}
are introduced and additional slow-roll conditions $|\epsilon_{s}| <1, |\eta_{s}| < 1$ are imposed. In this paper, however, being interested in the effect of the more general variation of sound velocity mentioned above, we consider the situation where these conditions are temporarily violated by the sudden change of the sound velocity.

In the following sections, we derive the approapriate variables for calculating the perturbation and the matching condition at the transition epoch. Then we evaluate the final power spectrum for two concrete examples of the sound velocity variation.

\section{Curvature perturbation and Basic equation}
\label{basic-eq}
The scalar perturbation from the Background FRW metric in the conformal Newtonian gauge is incorporated as 
\begin{equation}
ds^{2} = a^{2}\left[ -(1+2\Psi)d \tau^{2} + (1+2\Phi)\delta_{ij}dx^{i}dx^{j}\right] \label{eq:metric_pert}
\end{equation}
Denoting the inflaton perturbation $\delta \phi$ and the background value of the inflaton field $\bar{\phi}$, the most common perturbation variable, namely, comoving curvature perturbation $\zeta$ is defined as \cite{Mukhanov:1985rz,Mukhanov:1988jd,Sasaki:1986hm}
\begin{equation}
\zeta \equiv \Phi + \frac{H}{\dot{\bar{\phi}}}\delta\phi.
\end{equation}
The basic action and the equation of motion in the linear theory for $\zeta$ is written in terms of 
\begin{equation}
v=\zeta z,
\end{equation}
where a new parameter $z$ is defined by 
\begin{eqnarray}
z\equiv\frac{a\sqrt{2\epsilon}}{c_{s}}.
\end{eqnarray}
The second-order action for $v$ is derived by substituting (\ref{eq:metric_pert}) and $\phi=\bar{\phi}+\delta \phi$ into (\ref{eq:general_action}), neglecting the higher-order terms and using the background evolution equation. The result is
\begin{equation}
S^{(2)}=\frac{1}{2}\int d\tau d^{3}x \left(v^{\prime 2}+c_{s}^{2}v\Delta v+\frac{z^{\prime\prime}}{z}v^{2}\right),
\end{equation}
where a prime denotes a derivative with respect to the conformal 
time $\tau$. 
%which is given by 
%\begin{eqnarray}
%d\tau ={dt\over a(t)}.
%\end{eqnarray}
Then, the equation of motion for Fourier-transformed perturbation variable $v_{k}$ is given by
\begin{equation}
v_{k}^{\prime\prime}+\left(c_{s}^{2}k^{2}-\frac{z^{\prime\prime}}{z}\right)v_{k}=0.
\label{eq: basic eq-v}
\end{equation}

The above so-called Mukhanov-Sasaki equation (\ref{eq: basic eq-v}) is commonly used to discuss the behavior of the curvature perturbation. Here, however, we are interested in the time-dependence of the sound velocity, in which situation the potential term $z^{\prime\prime}/z$ in (\ref{eq: basic eq-v}) is not easy to treat since the variable $z$ also contains $c_{s}$. For example, if we consider a step-function-type variation of the sound velocity, a square term of $(c_{s}^{\prime}/c_{s})$ appears in $z^{\prime\prime}/z$, which becomes the square of the delta function $\delta(\tau-\tau_{0})$ and makes the analysis impossible where $\tau_{0}$ denotes time when sound velocity changes. 

It is therefore clear that the equation for $v_k$ is 
not suitable\footnote{As discussed in Appendix A, this is not the case in the curvaton scenario.} and so we should introduce a new variable $u_{k}$, which is related to $v_{k}$ as \cite{Mukhanov:text}
\begin{equation}
-c_{s}k^{2}u_{k}=z\left(\frac{v_{k}}{z}\right)^{\prime},
\ \ c_{s}v_{k}=\theta\left(\frac{u_{k}}{\theta}\right)^{\prime}, 
\label{eq: relation-u-v}
\end{equation}
where we have defined 
\begin{equation}
\theta\equiv\frac{1}{c_{s}z}.
\label{def: theta}
\end{equation}
The basic equation of motion (\ref{eq: basic eq-v}) for $v_{k}$ is translated into the new equation in terms of $u_k$, which is  
\begin{equation}
u_{k}^{\prime\prime}+\left(c_{s}^{2}k^{2}-\frac{\theta^{\prime\prime}}{\theta}\right)u_{k}=0.
\label{eq: basic eq-u}
\end{equation}
Note that the term $(c_{s}^{\prime}/c_{s})$ does not 
exist in $\theta''/\theta$ since the variable $\theta$ does not 
depend on $c_s$ due to the definition of $\theta$, (\ref{def: theta}). We have to solve this equation under the assumption that the background evolution satisfies the slow-roll conditions (\ref{eq:slow-roll_cond}).
 
The term $\theta''/\theta$ is rewritten in terms of slow-roll parameters as 
\begin{align}
\frac{\theta^{\prime\prime}}{\theta}=\frac{1}{\tau^{2}}\left(\frac{\eta}{2}+\epsilon\right),
\end{align}
where we have used slow-roll approximation (\ref{eq:slow-roll_cond}) and 
a useful equation 
\begin{equation}
aH=-\frac{1}{\tau(1-\epsilon)}. 
\end{equation}
Therefore, we obtain the basic equation as
\begin{equation}
u_{k}^{\prime\prime}+\left(c_{s}^{2}k^{2}-\frac{\nu^{2}-\frac{1}{4}}{\tau^{2}}\right)u_{k}=0,
\end{equation}
where we have defined 
\begin{equation}
\nu^{2} = \frac{\eta}{2}+\epsilon+\frac{1}{4}.
\end{equation}
and approximate it as 
\begin{equation}
\nu =\sqrt{\frac{\eta}{2}+\epsilon+\frac{1}{4}} \approx \frac{1}{2}+\frac{\eta}{2}+\epsilon.
\end{equation}

%---------  Matching ----------------------------------
\section{Step-like variation of the sound speed}
\label{step-variation}
In this section, we compute the curvature perturbation for a model such that time variation of the sound velocity is described by a step function as 
\begin{align}
c_{s}=
\begin{cases}
c_{s1} & (\tau < \tau_{0}) \\
c_{s2} & (\tau > \tau_{0})
\end{cases}.
\label{eq: varing c-s}
\end{align}
To take into account the transition of the sound velocity, it is important to impose a matching condition to the solution in (\ref{eq: basic eq-u}) at $\tau=\tau_0$ when the sound velocity suddenly changes. The matching condition is obtained by integrating (\ref{eq: basic eq-u}) in an infinitesimaly small time interval 
$[\tau_{0}-\delta \tau,\tau_{0}+\delta \tau]$, yielding two conditions;
\begin{equation}
u_{k}(\tau_{0}-\delta \tau)=u_{k}(\tau_{0}+\delta \tau),\ \ u_{k}^{\prime}(\tau_{0}-\delta \tau)=u_{k}^{\prime}(\tau_{0}+\delta \tau).
\label{eq:two_conditions}
\end{equation}
Hereafter, we use the following expression 
\begin{equation}
u_{k}(\tau_{0}-\delta \tau)\to u_{k1},\ \ u_{k}(\tau_{0}+\delta \tau)\to u_{k2}.
\end{equation}
In the regime when $\tau < \tau_{0}$, setting $c_{s}=c_{s1}$ leads to 
the equation of motion 
\begin{equation}
u_{k}^{\prime\prime}+\left(c_{s1}^{2}k^{2}-\frac{\nu^{2}-\frac{1}{4}}{\tau^{2}}\right)u_{k}=0,
\end{equation}
and its solution is obtained by 
\begin{equation}
u_{k1}=\sqrt{-kc_{s1}\tau}\left[d_{1}H_{\nu}^{(1)}(-kc_{s1}\tau)+d_{2}H_{\nu}^{(2)}(-kc_{s1}\tau)\right],
\end{equation}
where $H_{\nu}^{(1),(2)}(-kc_{s1}\tau)$ denote the Hankel functions and 
$d_{1,2}$ are constants to be 
determined by the initial condition at $\tau\to -\infty$. 
We choose the adiabatic vacuum at the initial time in terms of 
$v_k$: 
\begin{equation}
v_{k} \to \frac{1}{\sqrt{2kc_{s1}}}e^{-ikc_{s1}\tau}.
\end{equation}
From the equation (\ref{eq: relation-u-v}), we find
\begin{eqnarray}
-c_{s1}k^{2}u_{k1} = v_{k}^{\prime}, 
\end{eqnarray}
well inside the horizon, so we can take 
\begin{equation}
u_{k1} = \frac{i}{\sqrt{2c_{s1}}k^{3/2}}e^{-ikc_{s1}\tau}
\end{equation}
and match it with the limiting form of the Hankel function, 
\begin{equation}
H_{\nu}^{(1,2)}(x) \to \sqrt{\frac{2}{\pi x}}\exp\left[ i \left(\pm x-\frac{\nu\pi}{2}-\frac{\pi}{4}\right)\right], \ \ \ (x \to \infty),
\end{equation}
where the upper and lower signs correspond to the first and the second types, respectively.
Hence it leads to the choice of $d_1,d_2$ as
\begin{equation}
d_{1}=\frac{i}{2k^{3/2}}\sqrt{\frac{\pi}{c_{s1}}}\exp\left(\frac{2\nu+1}{4}\pi i\right), \ \ \ d_{2} =0.
\end{equation}
Neglecting all the phase factors which is irrelevant for calculating the power spectrum, the solution $u_{k1}$ takes the form 
\begin{equation}
u_{k1}=\frac{\sqrt{-\pi\tau}}{2k}H_{\nu}^{(1)}(-kc_{s1}\tau).
\end{equation}
Next, in the regime when $\tau>\tau_{0}$, setting $c_{s}=c_{s2}$ leads to 
\begin{align}
u_{k2} = \frac{\sqrt{-kc_{s2}\tau}}{2}\left[ \alpha_{k}^{\prime} H_{\nu}^{(1)}(-kc_{s2}\tau)+\beta_{k}^{\prime} H_{\nu}^{(2)}(-kc_{s2}\tau)\right],
\end{align}
and it is rewritten by 
\begin{align}
u_{k2} = \frac{\sqrt{-\pi\tau}}{2k}\left[ \alpha_{k} H_{\nu}^{(1)}(-kc_{s2}\tau)+\beta_{k} H_{\nu}^{(2)}(-kc_{s2}\tau)\right],
\label{sol: u-k2}
\end{align}
where we have defined
\begin{equation}
\alpha_{k} \equiv \sqrt{\frac{c_{s2}}{\pi}}k^{3/2}\alpha_{k}^{\prime},\ \ \ \beta_{k} \equiv \sqrt{\frac{c_{s2}}{\pi}}k^{3/2}\beta_{k}^{\prime}.
\end{equation}
From (\ref{eq:two_conditions}), we finally obtain the coefficients 
$\alpha_{k}$ and $\beta_{k}$ as 
\begin{align}
\alpha_{k} &= \frac{i\pi k\tau_{0}}{4}\left[c_{s2}H_{\nu}^{(1)}(-kc_{s1}\tau_{0})H_{\nu+1}^{(2)}(-kc_{s2}\tau_{0})-c_{s1}H_{\nu}^{(2)}(-kc_{s2}\tau_{0})H_{\nu+1}^{(1)}(-kc_{s1}\tau_{0})\right], \label{eq: sol-alpha} \\
\beta_{k} &= -\frac{i\pi k\tau_{0}}{4}\left[c_{s2}H_{\nu}^{(1)}(-kc_{s1}\tau_{0})H_{\nu+1}^{(1)}(-kc_{s2}\tau_{0})-c_{s1}H_{\nu}^{(1)}(-kc_{s2}\tau_{0})H_{\nu+1}^{(1)}(-kc_{s1}\tau_{0})\right],
\label{eq: sol-beta}
\end{align}
where we have used the following relation among the Hankel functions, 
\begin{eqnarray}
H_{\nu+1}^{(1)}(x)H_{\nu}^{(2)}(x)-H_{\nu+1}^{(2)}(x)H_{\nu}^{(1)}(x)
=-{4 i\over \pi x}.
\end{eqnarray}

%------------------------------------------------------

We have to estimate the power spectrum of $\zeta$ at the final time $\tau\to 0$. By using (\ref{eq: relation-u-v}), the curvature perturbation can be rewritten in terms of the variable $u$ as 
%\begin{align}
%\zeta &= \frac{v}{z}=\theta^{2}\left(\frac{u}{\theta}\right)^{\prime}=\theta\le%ft(u^{\prime}-\frac{\theta^{\prime}}{\theta} u\right) \nonumber \\
% &=\frac{1}{a\sqrt{2\epsilon}}\left(u^{\prime}-aH\left(1-\frac{\eta}{2}\right)u%\right).
%\label{eq: zeta-u relation}
%\end{align}
\begin{align}
\zeta_{k} &= \frac{v_{k}}{z}=\theta^{2}\left(\frac{u_{k}}{\theta}\right)^{\prime}=\theta\left(u_{k}^{\prime}-\frac{\theta^{\prime}}{\theta} u_{k}\right) \nonumber \\
 &\approx \frac{H}{\sqrt{2\epsilon}}\left(1-\frac{\eta}{2}\right)u_{k}.
\label{eq: zeta-u relation}
\end{align}
where, at the last transformation, we have chosen the super-horizon limit and neglected $u_{k}^{\prime}$ term.
%ThereforeAwe need the final values expressed as $u_{k2}(\tau\to 0)$ and 
%$u_{k2}^{\prime}(\tau\to 0)$. From the solution (\ref{sol: u-k2}), we obtain 
Therefore, we need the final values expressed as $u_{k2}(\tau\to 0)$. From the solution (\ref{sol: u-k2}), we obtain 
%\begin{align}
%u_{k2}^{\prime}(\tau \to 0) &\simeq \frac{\sqrt{-\pi\tau}}{2k}\frac{i}{\pi}\Gam%ma(\nu)\left(\frac{-kc_{s2}\tau}{2}\right)^{-\nu}\frac{1}{2\tau}(1-2\nu)(\alpha%^{\prime}-\beta^{\prime}), \nonumber\\
%u_{k2}(\tau \to 0) &\simeq \frac{\sqrt{-\pi\tau}}{2k}\frac{i}{\pi}\Gamma(\nu)\l%eft(\frac{-kc_{s2}\tau}{2}\right)^{-\nu}(\alpha^{\prime}-\beta^{\prime}),
%\label{eq: u-k2 limiting}
%\end{align}
\begin{align}
u_{k2}(\tau \to 0) &\simeq \frac{\sqrt{-\pi\tau}}{2k}\frac{i}{\pi}\Gamma(\nu)\left(\frac{-kc_{s2}\tau}{2}\right)^{-\nu}(\alpha_{k}-\beta_{k}),
\label{eq: u-k2 limiting}
\end{align}
where $\Gamma(x)$ denotes the Gamma function and we have used the limiting form of the Hunkel function as 
\begin{equation}
H_{\nu}^{(1)}(y)\simeq -H_{\nu}^{(2)}(y) \simeq \frac{i}{\pi}\Gamma(\nu)\left(\frac{y}{2}\right)^{-\nu}, \ \ \ (y \to 0).
\end{equation}
By using (\ref{eq: zeta-u relation}) and (\ref{eq: u-k2 limiting}), we can calculate the power spectrum of curvature perturbation as 
%\begin{eqnarray}
%P_{\zeta}(k)=|\zeta(\tau\to 0)|^{2}\approx 
%{4^\nu\Gamma^2(\nu)\over 8\pi k^2}  {H^2\over \epsilon}\left[aH(1-\epsilon)\rig%ht]^{2\nu-1} (kc_{s2})^{-2\nu} \left({25\over 16}+{5\epsilon\over 8}-{15\eta
%\over 8}\right)|\alpha'-\beta'|^2\bigg|_{aH=k c_{s2}}\hspace{-0.5cm}+O(\epsilon%^2, \eta^2),
%\end{eqnarray}
\begin{eqnarray}
P_{\zeta}(k)=|\zeta_{k}(\tau\to 0)|^{2}\approx 
{4^\nu\Gamma^2(\nu)\over 8\pi k^2}  {H^2\over \epsilon}\left[aH(1-\epsilon)\right]^{2\nu-1} (kc_{s2})^{-2\nu} \left(1+\eta\right)|\alpha_{k}-\beta_{k}|^{2}+\mathcal{O}(\epsilon^2, \eta^2),
\end{eqnarray}
where we have used the slow-roll approximation. 
Hence the term $|\alpha_{k}-\beta_{k}|^{2}$ is important to 
determine the power spectrum. $\alpha_{k}-\beta_{k}$ can be described by using (\ref{eq: sol-alpha}) and (\ref{eq: sol-beta}) as 
%\begin{align}
%\alpha^{\prime}-\beta^{\prime} &=\frac{i\pi k\tau_{0}}{4}\bigg[ c_{s2}H_{\nu}^{(1)}(kc_{s1}\tau_{0})\left(H_{\nu+1}^{(1)}(-kc_{s2}\tau_{0})+H_{\nu+1}^{(2)}(-kc_{s2}\tau_{0})\right) \nonumber \\
% &\ \ \ \ \ \ -c_{s1}H_{\nu+1}^{(1)}(-kc_{s1}\tau_{0})\left(H_{\nu}^{(1)}(-kc_{%s2}\tau_{0})+H_{\nu}^{(2)}(-kc_{s2}\tau_{0})\right)\bigg] \nonumber \\
% &= \frac{i\pi k\tau_{0}}{2}\biggl[ \Bigl(c_{s2}J_{\nu}(-kc_{s1}\tau_{0})J_{\nu%+1}(-kc_{s2}\tau_{0})-c_{s1}J_{\nu+1}(-kc_{s1}\tau_{0})J_{\nu}(-kc_{s2}\tau_{0}%)
%\Bigr) \nonumber \\
% &\ \ \ \ \ \ +i\Bigl(c_{s2}N_{\nu}(-kc_{s1}\tau_{0})J_{\nu+1}(-kc_{s2}\tau_{0}%)-c_{s1}N_{\nu+1}(-kc_{s1}\tau_{0})J_{\nu}(-kc_{s2}\tau_{0})\Bigr)\biggr],
%\end{align}
\begin{align}
\alpha_{k}-\beta_{k} &= \frac{i\pi k\tau_{0}}{2}\biggl[ \Bigl(c_{s2}J_{\nu}(-kc_{s1}\tau_{0})J_{\nu+1}(-kc_{s2}\tau_{0})-c_{s1}J_{\nu+1}(-kc_{s1}\tau_{0})J_{\nu}(-kc_{s2}\tau_{0})
\Bigr) \nonumber \\
 &\ \ \ \ \ \ +i\Bigl(c_{s2}N_{\nu}(-kc_{s1}\tau_{0})J_{\nu+1}(-kc_{s2}\tau_{0})-c_{s1}N_{\nu+1}(-kc_{s1}\tau_{0})J_{\nu}(-kc_{s2}\tau_{0})\Bigr)\biggr],
\end{align}
where $J_\nu,N_\nu$ are the Bessel functions with 
$\nu={1\over 2}+{\eta\over 2}+\epsilon$. Then we can calculate $|\alpha_{k}-\beta_{k}|^{2}$ as 
\begin{align}
|\alpha_{k}-\beta_{k}|^{2}&=\frac{\pi^{2} k^{2}\tau_{0}^{2}}{4}
\biggl[ \Bigl(c_{s2}J_{\nu}(-kc_{s1}\tau_{0})J_{\nu+1}(-kc_{s2}\tau_{0})-c_{s1}J_{\nu+1}(-kc_{s1}\tau_{0})J_{\nu}(-kc_{s2}\tau_{0})\Bigr)^{2} \nonumber \\
 &\ \ \ \ \ \ +\Bigl(c_{s2}N_{\nu}(-kc_{s1}\tau_{0})J_{\nu+1}(-kc_{s2}\tau_{0})-c_{s1}N_{\nu+1}(-kc_{s1}\tau_{0})J_{\nu}(-kc_{s2}\tau_{0})\Bigr)^{2}\biggr].
\end{align}
Here, we define new variables $A$ and $k_0$ as 
\begin{equation}
c_{s2}=Ac_{s1},\ \ -c_{s2}\tau_{0}=\frac{1}{k_{0}},\ \ -c_{s1}\tau_{0}=\frac{1}{Ak_{0}},
\end{equation}
and using them leads to 
\begin{align}
|\alpha_{k}-\beta_{k}|^{2}&=\frac{\pi^{2}k^{2}}{4k_{0}^{2}A^{2}}\bigg\{ A^{2}J_{\nu+1}^{2}\left(\frac{k}{k_{0}}\right)\left[J_{\nu}^{2}\left(\frac{k}{Ak_{0}}\right)+N_{\nu}^{2}\left(\frac{k}{Ak_{0}}\right)\right]+J_{\nu}^{2}\left(\frac{k}{k_{0}}\right)\left[J_{\nu+1}^{2}\left(\frac{k}{Ak_{0}}\right)+N_{\nu+1}^{2}\left(\frac{k}{Ak_{0}}\right)\right] \nonumber \\
 &\ \ \ \ \ \ -2AJ_{\nu+1}\left(\frac{k}{k_{0}}\right)J_{\nu}\left(\frac{k}{k_{0}}\right)\left[J_{\nu}\left(\frac{k}{Ak_{0}}\right)J_{\nu+1}\left(\frac{k}{Ak_{0}}\right)+N_{\nu}\left(\frac{k}{Ak_{0}}\right)N_{\nu+1}\left(\frac{k}{Ak_{0}}\right)\right]\bigg\}. 
\end{align}
Finally, we can evaluate the power spectrum at the 
sound horizon crossing $aH=k c_{s2}$ as 
\begin{eqnarray}
P_{\zeta}(k)={2^{1+\eta+2\epsilon}\over 8\pi} \Gamma^2\left({1\over 2}+{\eta\over 2}+\epsilon\right){(1-\epsilon)^{\eta+2\epsilon}\over \epsilon}  
{H^2\over k^3}\left(1+\eta\right)|\alpha_{k}-\beta_{k}|^2\bigg|_{aH=k c_{s2}}.
\label{eq:power}
\end{eqnarray}
%with 
%\begin{eqnarray}
%|\alpha'-\beta'|^2={\pi^2 A^2 x^2\over 4}\bigg[ A^{2}J_{\nu+1}^{2}\left(x^2\right)\left(J_{\nu}^{2}\left({x\over A}
%\right)+N_{\nu}^{2}\left({x\over A}\right)\right)
%+J_{\nu}^{2}\left(x\right)\left(J_{\nu+1}^{2}\left({x\over A}\right)
%+N_{\nu+1}^{2}\left({x\over A}\right)\right) \nonumber \\
%-2AJ_{\nu+1}\left(x\right)J_{\nu}\left(x\right)\left(
%J_{\nu}\left({x\over A}\right)J_{\nu+1}\left({x\over A}\right)
%+N_{\nu}\left({x\over A}\right)N_{\nu+1}\left({x\over A}
%\right)\right)\bigg],
%\end{eqnarray}
In the lowest-order slow-roll approxmation, we can set $\nu={1\over 2}$, in which case the Bessel functions are expressed by trigonometric functions and the modulation factor $|\alpha_{k}-\beta_{k}|^{2}$ can be recast to the following simple form (see Fig.1(a)). 
\begin{align}
|\alpha_{k}-\beta_{k}|^2=A\left[1+\left(\frac{1}{A^{2}}-1\right)\sin^{2}\left(\frac{k}{k_{0}}\right)\right]. \label{eq:mod_kaidan}
\end{align}
In this case, neglecting the slow-roll correction in the numerator, the dimensionless power spectrum (\ref{eq:power}) becomes
\begin{equation}
\Delta^{2}_{\zeta}(k) \equiv {k^{3}\over 2\pi^{2}}P_{\zeta}(k)={H^{2}A\over 8\pi^{2}\epsilon c_{s2}} \left[1+\left(\frac{1}{A^{2}}-1\right)\sin^{2}\left(\frac{k}{k_{0}}\right)\right]. \label{eq:nodimpower}
\end{equation} 
This result recovers the usual power spectrum with constant sound velocity if we take $A=1$. In the large-scale limit $k\to 0$, we can drop the oscillatory term and the final expression becomes 
\begin{equation}
\Delta^{2}_{\zeta}(k) = {H^{2}\over 8\pi^{2}\epsilon c_{s1}}, \label{eq:power_large}
\end{equation} 
which represents the power spectrum for modes which cross the horizon far before the transition time $\tau_{0}$.
 
\begin{figure}
\subfigure[]{%
\includegraphics[width=0.48\columnwidth]{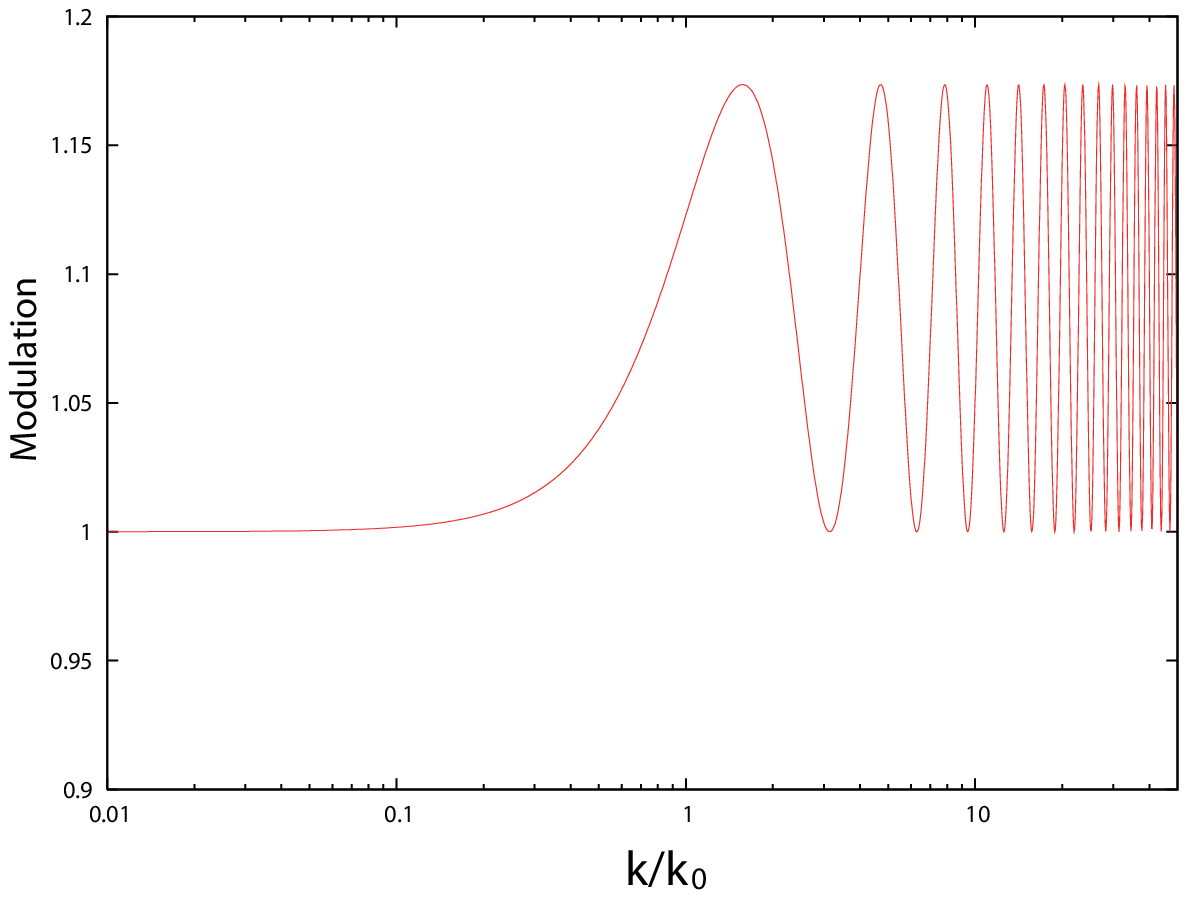}}%
\vspace{3mm}
\subfigure[]{%
\includegraphics[width=0.5\columnwidth]{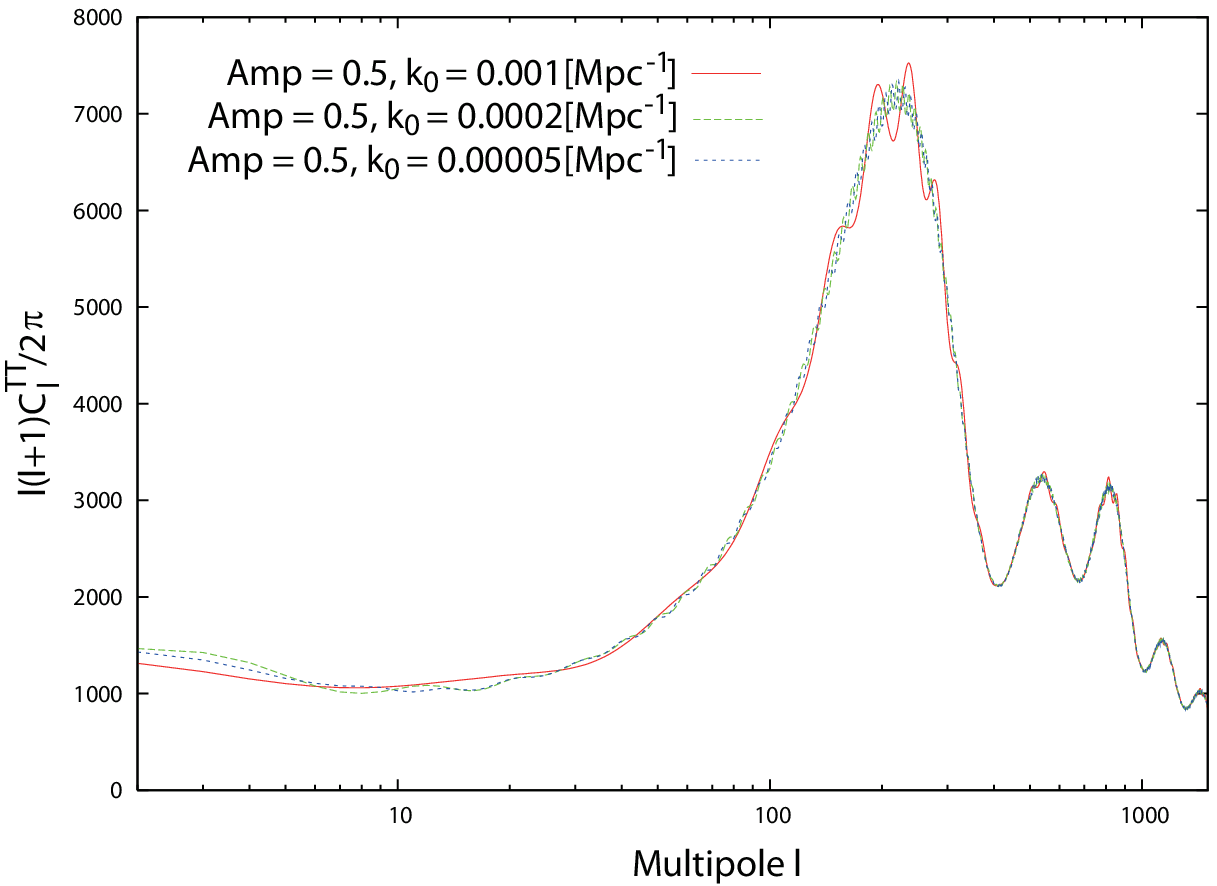}}%
\caption{(a) The modulation factor in the leading-order slow-roll approxmation (\ref{eq:mod_kaidan}) in the case $A=0.9$ ($A_{mp} = 0.17 $). (b) CMB temperature anisotropy spectrum for some different values of $k_{0}$. The other cosmological values are chosen as the WMAP 7-year mean values.}
\end{figure}

Taking the result (\ref{eq:nodimpower}) as the input primordial power specrum, or more quantitatively speaking, taking 
\begin{gather}
\Delta^{2}_{\zeta}(k) = A_{s} \left(\frac{k}{k_{piv}}\right)^{n_{s}-1} \left[1+A_{mp}\sin^{2}\left(\frac{k}{k_{0}}\right)\right], \label{eq:inputpower} \\
A_{s} = 2.43\times 10^{-9},\ k_{piv}=0.002[\mathrm{Mpc}^{-1}],\ n_{s}=0.963, \nonumber
\end{gather}
as the input one, we have computed CMB temperature anisotropy for several values of $k_{0}$ (see Fig.1(b)). Comparing (\ref{eq:inputpower}) with (\ref{eq:nodimpower}), we can clearly see that $A_{mp}\equiv 1/A^{2}-1$. The parameter $k_{0}$ determines the scale under which the rapid modulation appears. As we expect, if the value of $k_{0}$ is chosen as $\mathcal{O}(0.001[\mathrm{Mpc}^{-1}])$ or smaller, then the CMB spectrum starts to oscillate at the relatively large scale around $\ell \simeq 10$, which may better fit some anomalous data points obserbed in WMAP. Actually, as one exapmle value, if we take $k_{0}=0.003[\mathrm{Mpc}^{-1}]$, $A_{mp}=0.087$ and neglect the small scale ($\ell \ge 200)$ oscillation, we have found that the $\chi^{2}$-value improves $3.2$ compared with the usual case of power-law primordial power spectrum.

In Fig.2, we have plotted the difference between the CMB temperature anisotropy spectrum calculated by the primordial power spectrum with oscillations (\ref{eq:inputpower}) and the one by the usual power-law power spectrum without oscillations, which we denote $C_{\ell}^{osc}$ and $C_{\ell}^{std}$, respectively. Here the difference is defined as
\begin{equation}
\frac{\Delta C_{\ell}}{C_{\ell}} \equiv \frac{C_{\ell}^{osc}-C_{\ell}^{std}}{C_{\ell}^{std}}.
\end{equation}
Note that the overall amplitude $A_{s}$ in $C_{\ell}^{osc}$ is different from that used in $C_{\ell}^{std}$ by a factor $(1+A_{mp}/2)^{-1}$. We have also plotted the expected scatter of $C_{\ell}$ due to the cosmic variance, which is given by
%\footnote{The reason for this procedure is as follows. Here we want to know the difference purely due to the oscillation pattern, however the amplitude of the oscillating spectrum is amplified in small scale by the oscillation itself. Thus we need to correct this amplification originating from the oscillation when we make the corresponding standard spectrum without oscillation. $A_{mp}/2$ corresponds to the medium value of the amplitude of the oscillation.} The thick curve depicts 
\begin{equation}
\left(\frac{\Delta C_{\ell}}{C_{\ell}}\right)_{\mathrm{CV}} = \sqrt{\frac{2}{(2\ell+1)f_{sky}}},\ \ f_{sky}=0.65.
\end{equation} 
Note that the observational error of WMAP is cosmic-variance limited up to $\ell \simeq 400$ and it will be extended to $\ell \simeq 2500$  for PLANCK. Hence from this figure we can constrain $A_{mp}$ as $A_{mp} \lesssim 0.1$ now for $k_{0} = 10^{-4} [\mathrm{Mpc}^{-1}]$ which corresponds to the largest scale that can be measured with CMB experiments.

%This figure tells us that if the $k_{0}=0.0001[\mathrm{Mpc}^{-1}]$ which roughly corresponds to tha largest scale that can be measured by CMB experiment and the amplitude of the oscillation $A_{mp}$ is of order $\mathcal{O}(0.1)$ then, future CMB experiment such as PLANCK or Cosmic Variance limited experiment will be able to find those features.  

%\begin{figure}[htbp]
%\includegraphics[width=10cm]{a05k3way2.eps}
%\label{fig:cmb}
%\caption{CMB temperature anisotropy spectrum for some different values of $k_{0}$. The other cosmological values are chosen as the WAMP 7-year mean values.}
%\end{figure}

\begin{figure}[htbp]
\includegraphics[width=12cm]{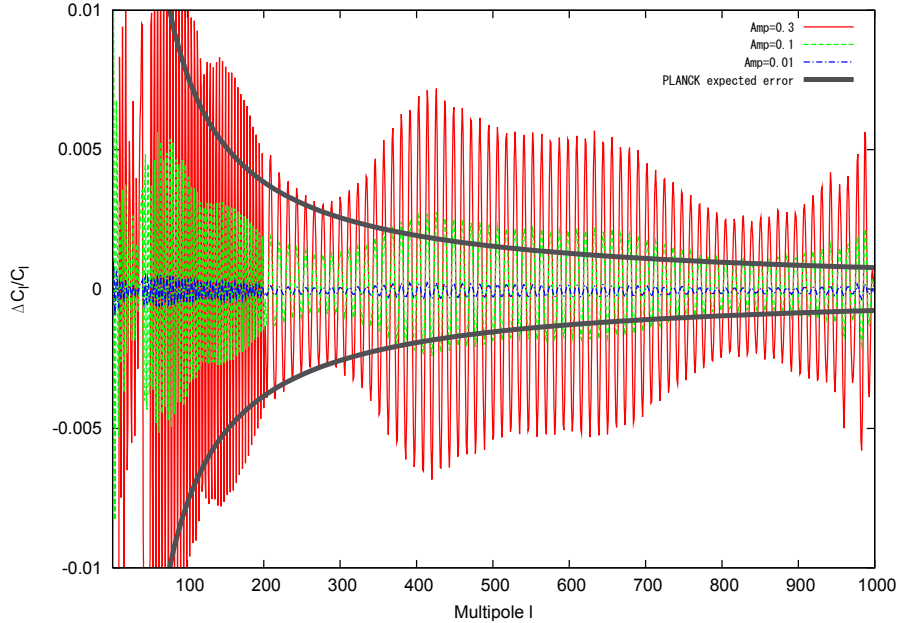}
\label{fig:cmb_diff}
\caption{The difference between CMB temperature anisotropy spectrum from the initial power spectrum with the oscillation and without the oscillation. We choose  $k_{0}=0.0001[\mathrm{Mpc}^{-1}]$ and the obserbational detection limit in the PLANCK-like experiment is also depicted.}
\end{figure}

This feature in the spectrum can be compared with the Trans-Planckian signatures \cite{Martin:2000xs,Martin:2000bv,Easther:2002xe}, with the particle production effect, or with the spectrum generated in the case the inflaton potential has a break in its slope. In the first case, the modification of the initial vacuum can change the relation and then leads to the ringing signatures in the primordial power spectrum. The exact form of the pattern depends on the choice of the hypersurface on which the initial condition is imposed. In the so-called Boundary Effective Field Theory (BEFT) approach \cite{Greene:2004np,Easther:2005yr}, the final expression becomes
\begin{equation}
\mathcal{P}_\mathrm{BEFT}(k)=A_{s}\left(\frac{k}{k_{0}}\right)^{n_{s}-1}\left[1+\frac{\beta k}{a_{i}\Lambda}\sin \left(2\frac{k}{a_{i}{H_{i}}}\right)\right]
\end{equation}
where $a_{i}$ and $H_{i}$ are the scale factor and the Hubble parameter on the initial condition hypersurface, respectively, and $\Lambda$ is a cutoff scale. This is similar to our result (\ref{eq:nodimpower}), but the amplitude of the oscillation depends on the wavenumber $k$, which makes crucial difference with our result.

In the second case, Barnaby et al. \cite{Barnaby:2009mc} discuss that the particle production (PP) make a bumplike contribution to the power spectrum, which is well fitted by 
\begin{equation}
\mathcal{P}_{\mathrm{PP}} (k) = A_{s}\left(\frac{k}{k_{0}}\right)^{n_{s}-1}+A_{\star}\left(\frac{k}{k_{\star}}\right)^{3}\exp\left(-\frac{\pi k^{2}}{2k_{\star}^{2}}\right).
\end{equation}
The feature is localized around $k_{\star}$, which makes sharp difference with our result.

In the last category, where the second derivative of the inflaton field has a sudden change at $\tau=\tau_{0}$, the modulation factor can be calculated similarly and the final expression is also similar to our result at the first glance. However, there is a big difference because in this case the slow-roll parameter has a step-like function over the transition time and the tilt of the spectrum takes the different values before and after the transition \cite{Joy:2007na,Joy:2008qd}.

\section{Top-hat type variation}
\label{top-hat-variation}
In this section, we consider the other type of variation of sound velocity, namely, the top-hat type. This is a simple extension of the model discussed in the last section. We parametrize the variation as
\begin{align}
c_{s}=
\begin{cases}
c_{s1} & (\tau < \tau_{0}) \\
c_{s2} & (\tau_{0} < \tau < \tau_{0}^{\prime}) \\
c_{s1} & (\tau_{0}^{\prime} < \tau)
\end{cases}.
\label{eq: varing c-s2}
\end{align}
and impose the same matching condition to the wavefunction $u_{k}$ as in (\ref{eq:two_conditions}) at the two transition times $\tau_{0}$ and $\tau_{0}^{\prime}$. The solutions of the equation of motion during constant sound velocity regimes are expressed in the same way as in the last section and we parametrize them as follows;
\begin{align}
u_{k1} &= P\frac{\sqrt{-\pi \tau}}{2k}H_{\nu}^{(1)}(-kc_{s1}\tau) \ \ (\tau < \tau_{0}), \\
u_{k2} &=\frac{\sqrt{-\pi\tau}}{2k}\left[ \alpha_{k1} H_{\nu}^{(1)}(-kc_{s2}\tau)+\beta_{k1} H_{\nu}^{(2)}(-kc_{s2}\tau)\right] \ \ (\tau_{0} < \tau < \tau_{0}^{\prime}), \label{eq:uk2} \\
u_{k3} &=\frac{\sqrt{-\pi\tau}}{2k}\left[ \alpha_{k2} H_{\nu}^{(1)}(-kc_{s1}\tau)+\beta_{k2} H_{\nu}^{(2)}(-kc_{s1}\tau)\right] \ \ (\tau_{0}^{\prime} < \tau), \label{eq:tophat_sol3}
\end{align}
where we have chosen the Bunch-Davies vacuum state at $\tau \ll \tau_{0}$. Two coefficients $\alpha_{k1}$ and $\beta_{k1}$ in (\ref{eq:uk2}) correspond to $\alpha_{k}$ and $\beta_{k}$ in (\ref{sol: u-k2}). Since the solutions and the matching conditions are the same, the coefficients $\alpha_{k1}$ and $\beta_{k1}$ are given by (\ref{eq: sol-alpha}) and (\ref{eq: sol-beta}). The matching conditions at $\tau=\tau_{0}^{\prime}$ gives the relations between $\alpha_{k2},\beta_{k2}$ and $\alpha_{k1},\beta_{k1}$;
\begin{align}
\alpha_{k2} &= -\frac{i\pi k\tau_{0}^{\prime}}{4}\Bigg[ \bigg\{c_{s2}H_{\nu+1}^{(1)}(-kc_{s2}\tau_{0}^{\prime})H_{\nu}^{(2)}(-kc_{s1}\tau_{0}^{\prime})-c_{s1}H_{\nu+1}^{(2)}(-kc_{s1}\tau_{0}^{\prime})H_{\nu}^{(1)}(-kc_{s2}\tau_{0}^{\prime})\bigg\}\alpha_{k1} \nonumber \\
           &\ \ \ \ +\bigg\{c_{s2}H_{\nu+1}^{(2)}(-kc_{s2}\tau_{0}^{\prime})H_{\nu}^{(2)}(-kc_{s1}\tau_{0}^{\prime})-c_{s1}H_{\nu+1}^{(2)}(-kc_{s1}\tau_{0}^{\prime})H_{\nu}^{(2)}(-kc_{s2}\tau_{0}^{\prime})\bigg\}\beta_{k1}\Bigg], \label{eq:alpha2} \\
\beta_{k2} &= \frac{i\pi k\tau_{0}^{\prime}}{4}\Bigg[\bigg\{ c_{s2}H_{\nu+1}^{(1)}(-kc_{s2}\tau_{0}^{\prime})H_{\nu}^{(1)}(-kc_{s1}\tau_{0}^{\prime})-c_{s1}H_{\nu+1}^{(1)}(-kc_{s1}\tau_{0}^{\prime})H_{\nu}^{(1)}(-kc_{s2}\tau_{0}^{\prime})\bigg\}\alpha_{k1} \nonumber \\
           &\ \ \ \ +\bigg\{c_{s2}H_{\nu+1}^{(1)}(-kc_{s2}\tau_{0}^{\prime})H_{\nu}^{(1)}(-kc_{s1}\tau_{0}^{\prime})-c_{s1}H_{\nu+1}^{(1)}(-kc_{s1}\tau_{0}^{\prime})H_{\nu}^{(1)}(-kc_{s2}\tau_{0}^{\prime})\bigg\}\beta_{k1}\Bigg]. \label{eq:beta2}
\end{align}

We want to know the power spectrum of the curvature perturbation at the final epoch $\tau\to 0$. As we have already seen, all we need is $u_{k3}(\tau\to 0)$. Combination of the solution (\ref{eq:tophat_sol3}) and the limiting form of the Hankel function leads to 
\begin{align}
u_{k3}(\tau \to 0) &\simeq \frac{\sqrt{-\pi\tau}}{2k}\frac{i}{\pi}\Gamma(\nu)\left(\frac{-kc_{s1}\tau}{2}\right)^{-\nu}(\alpha_{k2}-\beta_{k2}),
\end{align}
and the power spectrum of the curvature perturbation is described as
\begin{equation}
P_{\zeta}(k) \equiv |\zeta(\tau\to 0)|^{2} = \frac{H^{2}}{2\epsilon}\left(1+\frac{\eta}{2}\right)^{2}|u_{k3}(\tau\to 0)|^{2}. 
\end{equation} 
Also in this case, the factor to determine the modulation pattern is $|\alpha_{k2}-\beta_{k2}|^{2}$, which can be calculated by inserting (\ref{eq: sol-alpha}) and (\ref{eq: sol-beta}) into (\ref{eq:alpha2}) and (\ref{eq:beta2}). The final functional form however is so complicated. We therefore show the result only in the leading order slow-roll approximaton;
\begin{align}
|\alpha_{k2}-\beta_{k2}|^{2} &\simeq \cos^{2}\left[-kc_{s2}(\tau_{0}^{\prime}-\tau_{0})\right]+\sin^{2}\left[-kc_{s2}(\tau_{0}^{\prime}-\tau_{0})\right]\left[A^{2}\sin^{2}(-kc_{s1}\tau_{0}^{\prime})+\frac{1}{A^{2}}\cos^{2}(-kc_{s1}\tau_{0}^{\prime})\right] \nonumber \\
 &\ \ \ \ -2\left(A-\frac{1}{A}\right) \cos(-kc_{s1}\tau_{0}^{\prime})\sin(-kc_{s1}\tau_{0}^{\prime})\cos\left[-kc_{s2}(\tau_{0}^{\prime}-\tau_{0})\right]\sin\left[-kc_{s2}(\tau_{0}^{\prime}-\tau_{0})\right] + \mathcal{O}(\epsilon, \eta). \label{eq:mod2}
\end{align}

As expected, the modulation is determined by the combination of sine and cosine osillations. We can transform (\ref{eq:mod2}) to the following form.

\begin{align}
|\alpha_{k2}-\beta_{k2}|^{2}=&1-\frac{A^{2}-1}{2A}\bigg[(A^{2}-1)\sin^{2}\left[-kc_{s2}(\tau_{0}^{\prime}-\tau_{0})\right]-(A^{2}+1)\sin^{2}(-kc_{s1}\tau_{0}^{\prime}) \nonumber \\
 & + \frac{(A+1)^{2}}{2}\sin^{2}\left[-kc_{s1}\tau_{0}^{\prime}+kc_{s2}(\tau_{0}^{\prime}-\tau_{0})\right]+\frac{(A-1)^{2}}{2}\sin^{2}\left[-kc_{s1}\tau_{0}^{\prime}-kc_{s2}(\tau_{0}^{\prime}-\tau_{0})\right]\bigg]. \label{eq:mo2-2}
\end{align}

It is clear in this expression that there is no modulation when $A=1$ or $\tau_{0} = \tau_{0}^{\prime}$, which means no change in sound velocity. If we consider the long-wavelength limit ($k \to 0$), the dimensionless power spectrum of the curvature perturbation reduces to the same form as (\ref{eq:power_large})
%\begin{equation}
%\Delta_{\zeta}^{2}(k \to 0) \equiv {k^{3}\over 2\pi^{2}}P_{\zeta}(k \to 0)={H^{2} \over 8\pi^{2}\epsilon c_{s1}}, 
%\end{equation}
in the lowest-order slow-roll approximation of the numerator, which is valid because the long-wavelength mode exits the Hubble horizen when the sound velocity is $c_{s1}$ and superhorizen mode does not feel the sound velocity change.

\begin{figure}
\subfigure[]{%
\includegraphics[width=0.49\columnwidth]{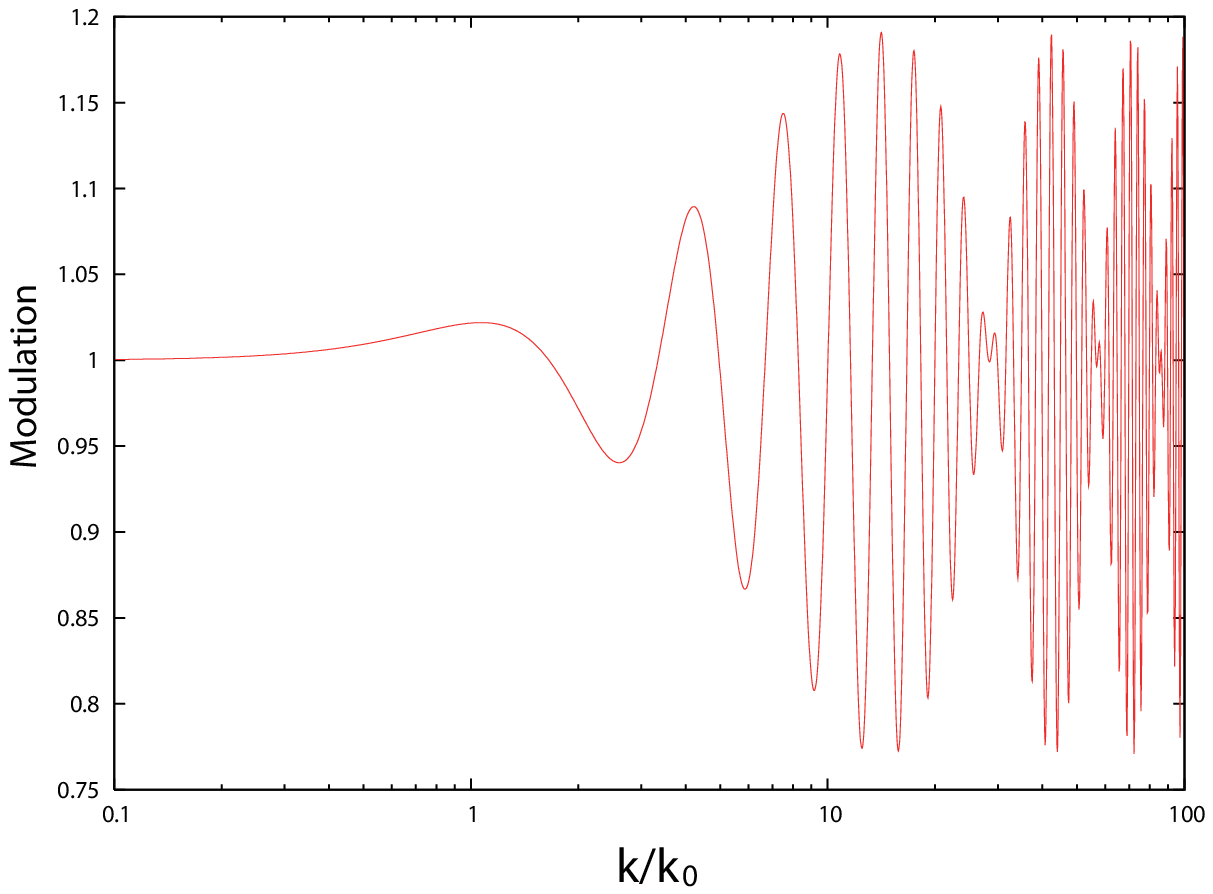}}%
\subfigure[]{%
\includegraphics[width=0.5\columnwidth]{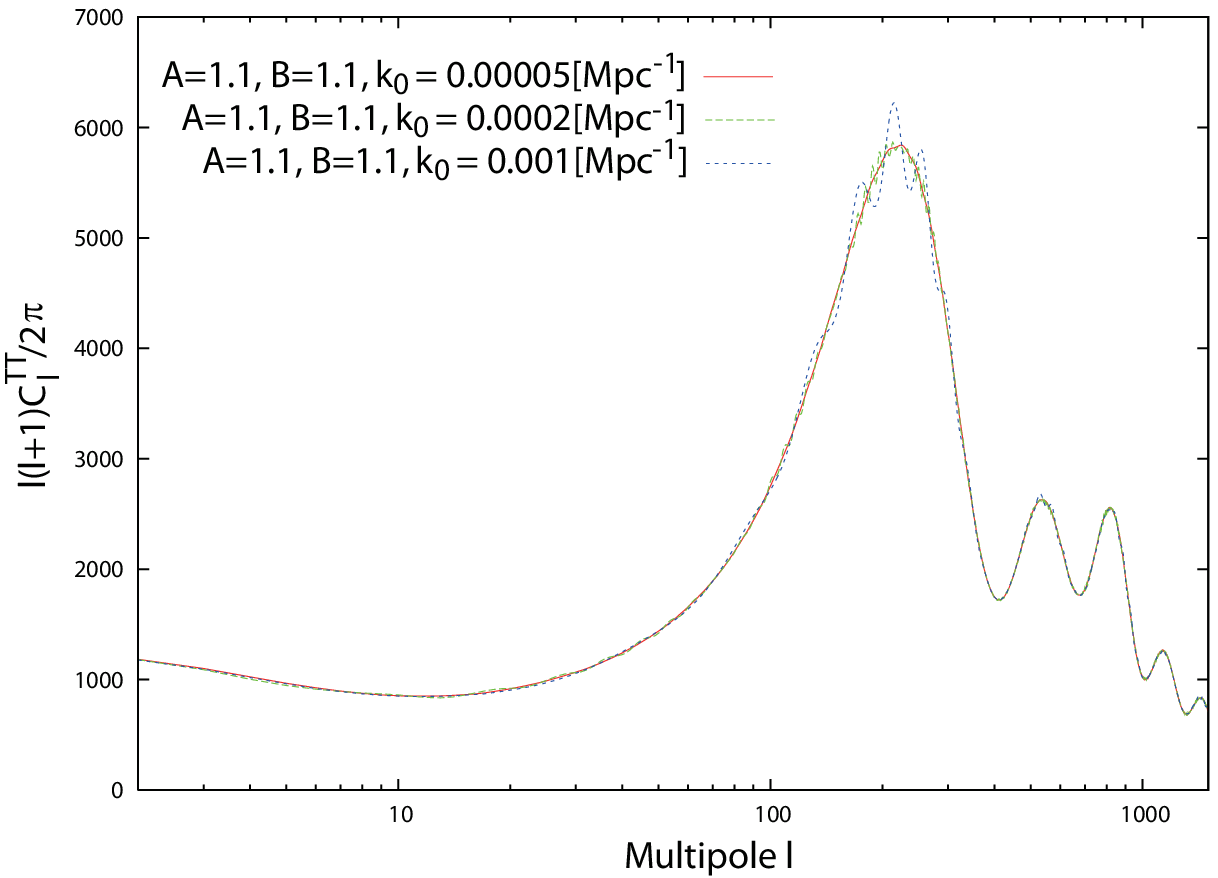}}%
\caption{(a) The modulation factor in the leading order slow-roll approxmation (\ref{eq:mo2-2}) in the case $A=1.1$ and $B=0.9$. (b) CMB temperature anisotropy spectrum for some different values of $k_{0}$. The other cosmological values are chosen as the WMAP 7-year mean values.}
\end{figure}

%\begin{figure}[htbp]
%\includegraphics[width=10cm]{beat_k32.eps}
%\label{fig:cmb2}
%\caption{CMB temperature anisotropy spectrum for some different values of $k_{0}$. The other cosmological values are chosen as the WAMP 7-year mean values.}
%\end{figure}

The modulation factor (\ref{eq:mod2}) or (\ref{eq:mo2-2}) is plotted in Fig.3(a) and the CMB temperature anisotropy spectrum is plotted in Fig.3(b). In both figures, we defined three parameters $k_{0}$, $A$ and $B$ as $-c_{s1}\tau_{0}^{\prime}=1/k_{0}$, $\tau_{0}=B\tau_{0}^{\prime}$ and $c_{s2} = A c_{s1}$, respectively. In these figures, We can see a beat, which is very natural because we now have two charactarizing parameters $c_{s1}\tau_{0}^{\prime}$ and $c_{s2}(\tau_{0}^{\prime}-\tau_{0})$ that determines the ocsillation periods and the final modulation pattern is the superposition of these waves. This result is easily extended to more general cases; if the variation of the sound velocity occurs suddenly and during it remains constant at other regimes, then the modulation of the power spectrum is some superpositions of several waves composed by trigonometric functions, which appears as a beat and the amplitude of the oscillation is independent of the wavenumber.

\section{Summary and discussion}
\label{sec:summary}

In summary, we have calculated the power spectrum of the curvature perturbation in case the sound velocity changes suddenly during inflation. In order to focus on the effects of the change of the sound velocity, we have assumed that the slow-roll parameters that reflect the background evolution are not affected and remains constant during the transition of the sound velocity and found that in the shorter-wavelength modes (with larger $k$) compared with the Hubble radius at the transition time there appears osillation patterns due to the mode mixing. In the simplest model in which the sound velocity experiences only one transition, the oscillation in the power spectrum is expressed as $\propto 1+A_{mp}\sin^{2} \left(\frac{k}{k_{0}}\right)$ where $k_{0}$ corresponds to the wavenumber that crosses the sound horizon at the transition time, $A_{mp}$ is defined as $A_{mp}= 1/A^{2}-1$ and $A$ is the ratio of sound velocity before and after the transition. Different from the trans-Planckian signatures or the models with a sudden change of the slow-roll parameters, the amplitude of the oscillation does not depend on the wavenumber, which gives a clue to find an evidence of the change of the sound velocity by some observations. Actually, we have computed the CMB temperature anisotropy using the primordial power spectrum we have obtained and we can see the oscillating patterns in the resulting angular power spectrum originating in the modulation of the primordial curvature perturbations. 

These oscillatory behaviors make it difficult to constrain the amplitude of the modulation using Markov Chain Monte Carlo (MCMC) analysis due to the problem of the convergence. Thus, we calculated the difference between the CMB temperature spectrum from the oscillatory primordial power spectrum and the one without the oscillation and compared it with the scatter due to the cosmic variance. We have confirmed that if $k_{0}=10^{-4}[\mathrm{Mpc^{-1}}]$ the current CMB experiment is sensitive to $A_{mp} \gtrsim 0.1$ or in other words is sensitive to $A\lesssim 0.95$. 

%Note that we have only calculated in the case $A \ge 1$ or $A_{mp} \ge 0$ for simplicity, however we can do the same calculation and get similar constraints for $A \le 1$ or $A_{mp} \le 0$.

%concluded that if the amplitude of the oscillation is of order $\mathcal{O}(0.1)$ then future CMB experiment such as PLANCK will find those features.

\begin{table}
\begin{tabular}{c||c|c|c|}
 & A=0.995 & A=0.99 & A=0.97 \\
 & ($A_{mp}=0.0099$)  & ($A_{mp}=0.0197$) & ($A_{mp}=0.0574$) \\
\hline \hline
$k_{0}=0.0001$ & -0.054 & -0.11 & -0.27 \\
\hline
$k_{0}=0.0003$ & -0.063 & -0.11 & -0.13 \\
\hline
$k_{0}=0.001$ & -0.34 & 0.60 & -0.81 \\
\hline
$k_{0}=0.003$ & 0.045 & 0.30 & 3.34 \\
\hline
\end{tabular}
\caption{This table shows the $\Delta \chi^{2}$ values in the step-function model described in Sec.\ref{step-variation} compared with WMAP 7-year best fit cosmological parameters in the simple $\Lambda$CDM model. The unit of the wavenumber $k_{0}$ is Mpc$^{-1}$. For the calculation of the $\chi^{2}$ values, we exploited the WMAP-7year likelihood funcion \cite{Larson:2010gs}. }
\label{table:deltachi1}
\end{table}

\begin{table}
\begin{tabular}{c||c|c|c|}
 & $A=1.005$ & $A =1.01$ & $A =1.03$ \\
\hline \hline
$k_{0}^{\prime}=0.00001$ & -2.26 & -1.54 & 2.53 \\
\hline
$k_{0}^{\prime}=0.00005$ & -1.88 & -1.40 & 1.28 \\
\hline
$k_{0}^{\prime}=0.0001$ & -1.72 & -1.25 & 1.33 \\
\hline
$k_{0}^{\prime}=0.0005$ & -2.01 & -1.59 & 2.28 \\
\hline
\end{tabular}
\caption{This table shows the $\Delta \chi^{2}$ values in the top-hat function model described in Sec.\ref{top-hat-variation} compared with WMAP 7-year best fit cosmological parameters in the simple $\Lambda$CDM model. We have defined a new variable $k_{0}^{\prime}$ as $-c_{s1}\tau_{0}\equiv 1/k_{0}^{\prime}$. The unit of the wavenumber $k_{0}^{\prime}$ is Mpc$^{-1}$. Here, we fix $k_{0}=0.001[\mathrm{Mpc}^{-1}]$. For the calculation of the $\chi^{2}$ values, we exploited the WMAP-7year likelihood funcion \cite{Larson:2010gs}. }
\label{table:deltachi2}
\end{table}

To discuss how WMAP experiment can constrain or find the amplitude of the oscillation of the primordial power spectrum, in Tables I and II, we have computed $\Delta \chi^{2}$ for various parameter values in the two models we have considered. We can check from this table that in the step-function model, too small a transition scale (or too large a transition wavenumber $k_{0}$) and too large an oscillation amplitude cannot fit the present CMB anisotropy data. In the top-hat function model, we have found several good improvements of $\Delta \chi^{2}$ values, though we have more parameters than step-function model in thic model. Although we have been unable to find parameter values that satisfy Akaike's Information Criterion \cite{Akaike}, several negative $\Delta \chi^{2}$ values may give us some motivations to search for the parameter values for much better fit to WMAP data or more accurate future CMB experiment data in our model.

%Basically, it is difficult to realize such a sudden transition of the sound velocity without violating the background slow-roll conditions. However, curvaton scenario or some special couplings between the inflaton field and some other fields give us chances to make conclete examples. Furthermore, these additional fields other than inflaton field has freedom to make perturbation power spectrum with features more naturally. We will look for these possibilities in the near future. 

%In the future work we want to study more quantitatively whether the effect of the sudden change of the sound velocity can be constrained or discovered from today's or future obserbations of CMB or Large Scale Structure surveys.  
 
%In this work, we have only considered the sudden transition model as the time-dependence of the sound velocity. Acturally, contrary to our models, there are several attempts to uncover the effect of time-varying sound velocity on the curvature perturbation. However they have only treated the limiting case: the very slow variation of sound velocity. In the near future, we will study more general case of the sound velocity variation and whether it can explain the several large-scale anomalies discovered from the WMAP data. 

%%%%%%%%%%%%%%%%%%%%%%%%%%%%%
\acknowledgments
M.N and R.S are supported by JSPS through research fellowships. 
The work of Y.T is based on 
the financial support by Research Center of the Early Universe (RESCEU), 
University of Tokyo and by JSPS Grant-in-Aid 
for Young Scientists (B) No. ~21740192. 
The work of J.Y is supported by JSPS Grant-in-Aid for Scientific Research No. 19340054 and Global COE Program ``The Physical Sciences Frontier", MEXT, Japan.

\section*{Appendix A}
In this appendix, we comment on the consistent background evolution which keeps slow-roll condition during the sudden transition of the sound velocity. In general, the slow-roll conditions are violated by changing the sound velocity so rapidly like step-type funtion without numerical fine-tuning between non-canonical kinetic terms.

One way to overcome the difficulty is to consider the curvaton scenario \cite{Moroi:2001ct,Lyth:2001nq}. If the curvaton Lagrangian has the following form,
\begin{align}
\mathcal{L} = \frac{1}{2}\partial_{\mu}\phi\partial^{\mu}\phi+\frac{1}{2}f(\phi)\partial_{\mu}\sigma\partial^{\mu}\sigma &+\frac{1}{4}g_{1}(\phi)(\partial_{\mu}\phi\partial^{\mu}\phi)^{2}+ \frac{1}{4}g_{2}(\phi)(\partial_{\mu}\sigma\partial^{\mu}\sigma)^{2}\nonumber \\
 &+\frac{1}{4}g_{3}(\phi)(\partial_{\mu}\phi\partial^{\mu}\phi)(\partial_{\nu}\sigma\partial^{\nu}\sigma)+\frac{1}{4}g_{4}(\phi)(\partial_{\mu}\phi\partial^{\mu}\sigma)^{2}+\cdots -V(\phi)-\frac{m^{2}}{2}\sigma^{2},  
\end{align}  
where $\phi$ and $\sigma$ denote the inflaton and the curvaton, respectively, and $g_{i}(\phi) = \mathcal{O}(\Lambda^{-4})$. Assuming that, during inflation, the curvaton is set to some constant value, we can write down the second-order Lagrangian of the curvaton perturbation
\begin{equation}
\mathcal{L}^{(2)}_{\sigma}=\frac{F(\phi)^{2}}{2a^{2}}\sigma^{\prime 2}-\frac{F(\phi)^{2}+G(\phi)}{2a^{2}}(\nabla\sigma)^{2}-\frac{m^{2}}{2}\sigma^{2}.
\end{equation}
Here,
\begin{align}
F(\phi)^{2} &= f(\phi)+\left(g_{3}(\phi)+g_{4}(\phi)\right)X_{\phi}+\cdots, \\
G(\phi) &= -g_{4}(\phi)X_{\phi} + \cdots,
\end{align}
where $X_{\phi}=\dot{\phi}^{2}/2$. Then we can read off the sound velocity for the curvaton perturbation as 
\begin{equation}
c_{s}^{2}=\frac{F(\phi)^{2}+G(\phi)}{F(\phi)^{2}}=\frac{f(\phi)+g_{3}(\phi)X_{\phi}}{f(\phi)+(g_{3}(\phi)+g_{4}(\phi))X_{\phi}}. \label{sound_curvaton}
\end{equation}
In order for $c_{s}$ to deviate from unity, there should be difference between the coefficients of time-derivative and of space-derivative. During inflation, the background breaks the symmetry between time and space through $\dot{\phi}$. Because of the coupling $\partial_{\mu}\phi\partial^{\mu}\sigma$, the sound velocity of $\sigma$ can be affected by $\dot{\phi}$. Actually the important factor to cause the deviation from unity in (\ref{sound_curvaton}) is $g_{4}(\phi)$, namely, the coefficient of the coupling $\partial_{\mu}\phi\partial^{\mu}\sigma$. This observation also implies that it is difficult to change the sound velocity without affecting background evolution in generic single field inflation models since, in such models, all the higher order couplings contribute to both time and space derivatives of the inflaton perturbation and hence the deviation in the sound speed is related to a form of a background equation of motion.

To be consistent with the assumption that the background value of the curvaton keeps constant, the condition 
\begin{equation}
\frac{\dot{\sigma}}{H\sigma}=\frac{1}{\sigma}\frac{{\rm d}\sigma}{{\rm d} N} \ll 1 \label{bg-condition}
\end{equation}
should be satisfied, where $N$ stands for the number of e-folds and an over dot denotes differentiation with respect to the physical time. The background equation of motion for $\sigma$ is 
\begin{equation}
\frac{{\rm d}^{2}\sigma}{{\rm d} N^{2}} +(3-\epsilon+2\alpha)\frac{{\rm d}\sigma}{{\rm d} N}+\left(\frac{m}{HF}\right)^{2}\sigma =0,
\end{equation}
where $\alpha\equiv\dot{F}/{HF}$. Imposing $\frac{1}{\sigma}\frac{{\rm d}^{2}\sigma}{{\rm d} N^{2}} \ll 1$, this equation tells
\begin{equation}
\frac{{\rm d}\sigma}{{\rm d}N}\simeq -\frac{1}{3-\epsilon+2\alpha}\left(\frac{m}{HF}\right)^{2}\sigma.
\end{equation}
Thus the condition (\ref{bg-condition}) is satisfied if $m/H \ll F$.

As for the curvaton perturbation, the second order Lagrangian can be written as
\begin{equation}
\sqrt{-g}\mathcal{L}^{(2)}_{\sigma}=a^{2}F^{2}\left[ \frac{1}{2}\sigma^{\prime 2}-\frac{1}{2}c_{s}^{2}(\nabla\sigma)^{2}-\frac{m^{2}a^{2}}{2F^{2}}\sigma^{2}\right]. \label{curvaton_pert_second}
\end{equation}
Introducing the new variable $v_{\sigma}\equiv z_{\sigma}\sigma$ where $z_{\sigma}\equiv aF$ which corresponds to $v$ for the inflaton scenario in the main text, (\ref{curvaton_pert_second}) can be rewritten as 
\begin{equation}
\sqrt{-g}\mathcal{L}^{(2)}_{\sigma}=\frac{1}{2}v_{\sigma}^{\prime 2}-\frac{1}{2}c_{s}^{2}(\nabla v_{\sigma})^{2}-\left(\frac{m^{2}a^{2}}{2F^{2}}-\frac{z_{\sigma}^{\prime\prime}}{z_{\sigma}}\right)v_{\sigma}^{2}.
\end{equation}
Now we can derive the evolution equation for the curvaton perturbation in Fourier space as 
\begin{equation}
v_{\sigma k}^{\prime\prime}+\left(c_{s}^{2}k^{2}+\frac{m^{2}a^{2}}{2F^{2}}-\frac{z_{\sigma}^{\prime\prime}}{z_{\sigma}}\right)v_{\sigma k}=0.
\end{equation}
Imposing the above-derived condition $m/H \ll F$, we can neglect the mass term in this equation during slow-roll inflation. On the other hand, the second derivative of $z_{\sigma}$ can be evalutated as 
\begin{equation}
\frac{z_{\sigma}^{\prime\prime}}{z_{\sigma}} = (aH)^{2}\left[ (1+\alpha)(2-\epsilon+\alpha)+\frac{{\rm d}\alpha}{{\rm d}N}\right].
\end{equation}
If we want to change the value of sound velocity without affecting the background evolution, $F(\phi)$ should be fixed. This is achieved when $f(\phi)=1$ and $g_{3}(\phi)=-g_{4}(\phi) =g(\phi)$. In this case, $c_{s}^{2}=1+g(\phi)X_{\phi}+\cdots$ and we can change only $c_{s}$ by choosing $g(\phi)$ as an appropriate form with $F=1$. Actually, now that $z_{\sigma}^{\prime\prime}/z_{\sigma} =(aH)^{2}(2-\epsilon)$, the techniques to calculate the power spectrum introduced in the main text can be easily applied without introducing the new perturbation variable $u_{k}$ which is no longer convinient to use.

%which can be changed relatively freely by choosing the coupling funtion $f(\phi)$, $g_{1}(\phi)$ and $g(\phi)$ without violating the slow-roll condition for the inflaton field. Of course, not to affect the background inflaton motion from the curvaton motion, we need to tune the relation between these coupling functions.

%In the curvaton case, the perturbation variable $u_{\sigma k}$ is no longer convinient to use since the curvaton is subdominant during inflation and only contributes to the perturbations, in other words, the curvaton does not become the source of the gravitational potential which the new variable $u_{\sigma k}$ represents. However, in this case, we can treat the evolution equation for the original variable $v_{\sigma k}$, and then we can get the same power spectrum with the one obtained in the main part of this paper.

\end{document}